\documentclass[preprint,aps,showpacs,nofootinbib,preprintnumbers,amsmath,amssymb]{revtex4-1}
\usepackage{}
\usepackage{epsfig}
\usepackage{subfigure}
\usepackage{dcolumn}
\usepackage{bm}
\usepackage[usenames ,dvipsnames]{xcolor}
\usepackage{slashed}
\usepackage{graphicx,color}

\begin{document}
\title{Direct CP violation in $\Lambda_b$ decays}

\author{Y.K. Hsiao$^{1,2}$ and C.Q. Geng$^{1,2,3}$}
\affiliation{
$^{1}$Physics Division, National Center for Theoretical Sciences, Hsinchu, Taiwan 300\\
$^{2}$Department of Physics, National Tsing Hua University, Hsinchu, Taiwan 300\\
$^{3}$Chongqing University of Posts \& Telecommunications, Chongqing, 400065, China
}
\date{\today}

\begin{abstract}
We study direct CP violating asymmetries (CPAs) in the two-body $\Lambda_b$ decays of $\Lambda_b\to pM(V)$
with $M(V)=K^-(K^{*-})$ and $\pi^-(\rho^-)$ based on the generalized factorization method.
After simultaneously explaining the observed decay branching ratios of
$\Lambda_b\to (p K^-\,,\; p \pi^-)$ with  ${\cal R}_{\pi K}\equiv {\cal B}(\Lambda_b\to p \pi^-)/{\cal B}(\Lambda_b\to p K^-)$
being $0.84\pm 0.09$, we find  that their corresponding direct CPAs are $(5.8\pm 0.2,\,-3.9\pm 0.2)\%$ in the standard model (SM),
in comparison with $(-5^{+26}_{-~5},\, -31^{+43}_{-~1})\%$ and $(-10\pm8\pm4,\, 6\pm7\pm3)\%$ from the perturbative QCD
calculation and the CDF experiment, respectively. For $\Lambda_b\to ( p K^{*-},\, p\rho^-)$, the  decay branching ratios and
CPVs in the SM are predicted to be $(2.5\pm0.5,\,11.4\pm2.1)\times 10^{-6}$ with ${\cal R}_{\rho K^*}=4.6\pm0.5$ and
$(19.6\pm1.6,\, -3.7\pm0.3)\%$, respectively. The uncertainties for the CPAs in these decay modes as well as
${\cal R}_{\pi K,\,\rho K^{*-}}$ mainly arise from the quark mixing elements and non-factorizable effects,
whereas those from the hadronic matrix elements are either totally eliminated or small. We point out that the large CPA
for $\Lambda_b\to p K^{*-}$ is promising to be measured by the CDF and LHCb experiments,
which is a clean test of the SM.
\end{abstract}

\maketitle
\section{introduction}
It is known that one of the main goals in the $B$ meson
factories is to confirm the weak CP phase in the
Cabbibo-Kobayashi-Maskawa (CKM) paradigm~\cite{CKM} of 
the Standard Model (SM)
through CP violating effects. 
Needless to say that the origin of CP violation is the most fundamental 
problem in physics, which may also shed light on the puzzle of the 
matter-antimatter asymmetry in the Universe.
However, the direct CP violating
asymmetries (CPAs), ${\cal A}_{CP}$, in $B$ decays have not been
clearly understood yet. In particular,  the naive result of ${\cal
A}_{CP}(\bar B^0\to K^-\pi^+)\simeq {\cal A}_{CP}(B^-\to K^-\pi^0)$
in the SM, cannot be approved by the experiments~\cite{Li_Kpi}.
It is known that it is inadequate to calculate the direct CPAs in the
two-body mesonic $B$ decays
due to the limited  knowledges on strong phases~\cite{Hou:2006jy}.
Clearly, one should look for CPV effects in other processes, in which the hadronic
effects are well understood.

Unlike the two-body $B$ meson decays, due to the flavor conservation,
there is neither color-suppressed  nor annihilation contribution
in the two-body baryonic modes of
$\Lambda_b\to p K^-$ and $\Lambda_b \to p \pi^-$,
providing  the controllable nonfactorizable effects and
traceable strong phases for the CPAs.
In fact, their decay branching ratios have been recently observed,
given by~\cite{pdg}
\begin{eqnarray}\label{exbr}
{\cal B}(\Lambda_b\to p K^-)&=&(4.9\pm 0.9)\times 10^{-6}\,,\nonumber\\
{\cal B}(\Lambda_b\to p \pi^-)&=&(4.1\pm 0.8)\times 10^{-6}\,.
\end{eqnarray}
Although the two decays have been extensivily discussed
in the leterature~\cite{Lu:2009cm,Wei:2009np,Wang:2013upa},
the measured values  in Eq.~(\ref{exbr}) cannot be
simultaneously explained in the studies.

In this paper, we will first examine
these two-body baryonic decays based on
the configuration of
the $\Lambda_b\to p$ transition with a recoiled $K$ or $\pi$, and then calculate
${\cal A}_{CP}(\Lambda_b\to p K^-,p\pi^-)$, which have been measured by
the CDF collaboration~\cite{Aaltonen}.
We will also extend our study to the corresponding vector modes of $\Lambda_b\to p V$ with $V=K^{*-}(\rho^-)$
as well as other two-body beauty baryons (${\cal B}_b$) decays, such as $\Xi_b$, $\Sigma_b$ and $\Omega_b$.

\section{Formalism}
\begin{figure}[t!]
\centering
\includegraphics[width=2.5in]{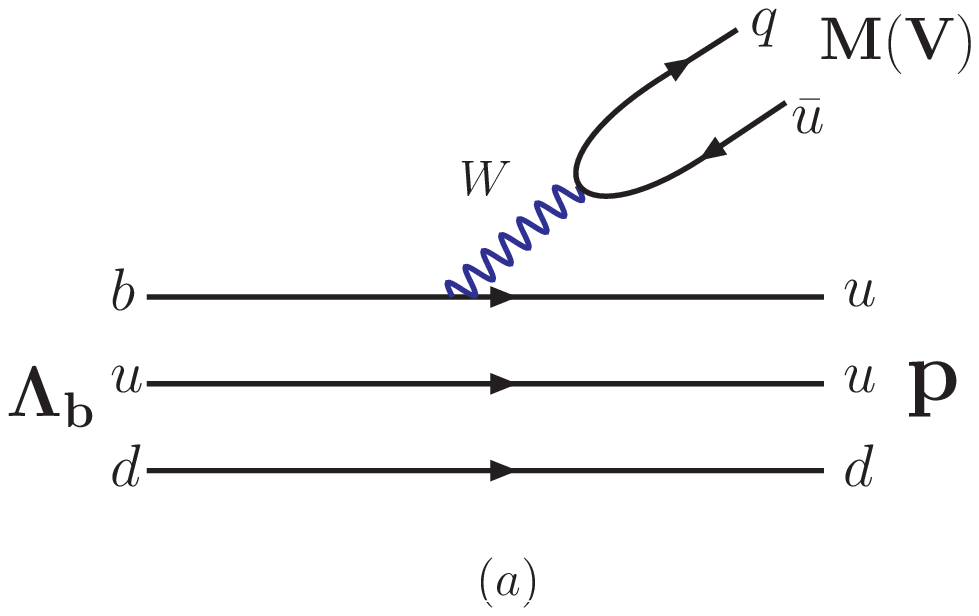}
\includegraphics[width=2.5in]{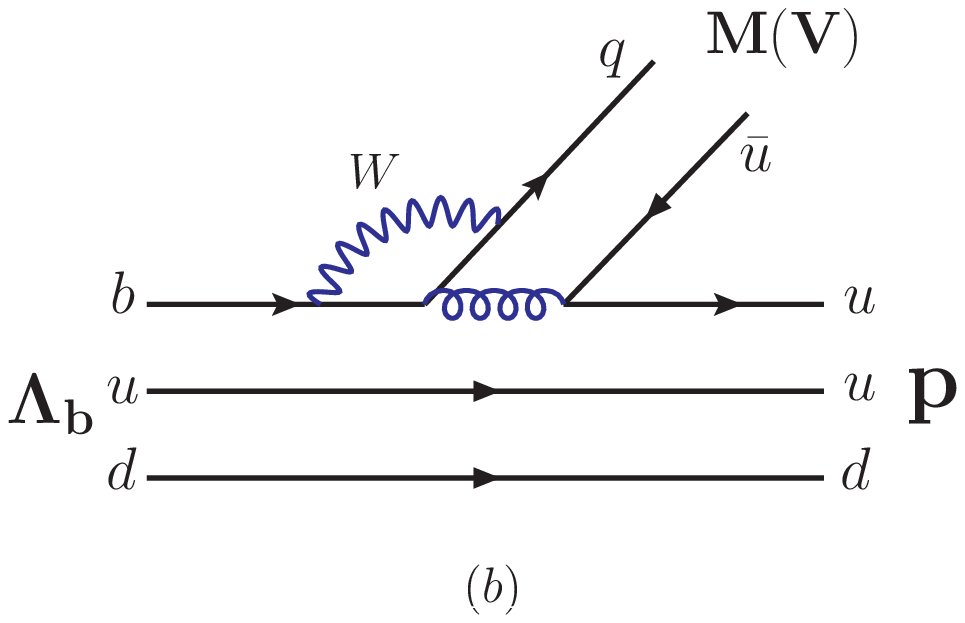}
\caption{Contributions to $\Lambda_b\to pM(V)$ from
(a)  color-allowed tree-level  and
(b)  penguin diagrams.}\label{LbtopM}
\end{figure}
According to the decaying processes depicted in Fig.~\ref{LbtopM},
in the generalized factorization approach~\cite{ali}
the amplitudes of $\Lambda_b\to p M(V)$ with $M(V)=K^-(K^{*-})$ and $\pi^-(\rho^-)$
can be derived as
\begin{eqnarray}\label{eq1}
{\cal A}(\Lambda_b\to p M)&=&i\frac{G_F}{\sqrt 2}m_b f_M\bigg[
\alpha_{M}\langle p|\bar u b|\Lambda_b\rangle+
\beta_{M}\langle p|\bar u\gamma_5 b|\Lambda_b\rangle\bigg]\,, \nonumber\\
{\cal A}(\Lambda_b\to p V)&=&\frac{G_F}{\sqrt2}m_{V}f_{V}
\varepsilon^{\mu*}\alpha_{V}\langle p|\bar u\gamma_\mu(1-\gamma_5) b|\Lambda_b\rangle\;,
\end{eqnarray}
 where $G_F$ is the Fermi constant and the meson decay constants $f_{M(V)}$ are defined by
 $\langle M|\bar q_1\gamma_\mu \gamma_5 q_2|0\rangle=-if_M q_\mu~$
and $\langle V|\bar q_1\gamma_\mu q_2|0\rangle=m_{V} f_{V}\varepsilon_\mu^*$
with
the four-momentum $q_\mu$ and polarization $\varepsilon_\mu^*$, respectively.
The constants $\alpha_{M}$ ($\beta_M$) and $\alpha_{V}$ in Eq.~(\ref{eq1}) are related to
the (pseudo)scalar and vector or axialvector quark currents,  given by
\begin{eqnarray}\label{eq2}
\alpha_{M}(\beta_{M})&=& V_{ub}V_{uq}^*a_1-V_{tb}V_{tq}^*(a_4\pm r_M a_6)\;,\nonumber\\
\alpha_{V}&=& V_{ub}V_{uq}^*a_1-V_{tb}V_{tq}^*a_4\;,
\end{eqnarray}
where $r_M\equiv {2 m_M^2}/[m_b (m_q+m_u)]$,
$V_{ij}$ are the CKM matrix elements,  $q=s$ or $d$,
and  $a_i\equiv c^{eff}_i+c^{eff}_{i\pm1}/N_c^{(eff)}$ for $i=$odd (even)
are composed of the effective Wilson coefficients $c_i^{eff}$ defined in Ref.~\cite{ali}.
We note that, as seen from Fig.~\ref{LbtopM},
there is no annihilation diagram at the penguin level
for $\Lambda_b\to p M(V)$, unlike
the cases in the two-body mesonic $B$ decays.
In addition,
without the color-suppressed tree-level diagram,
the non-factorizable effects in these baryonic decays can be modest.
In order to take account of
the non-factorizable effects, we use the generalized factorization method by setting
 the color number as $N_c^{eff}$,
which floats from 2 to $\infty$.
The matrix elements of the ${\cal B}_b\to {\cal B}$ baryon
transition in Eq. (\ref{eq1}) have the general forms:
\begin{eqnarray}
&&\langle {\cal B}|\bar q \gamma_\mu b|{\cal B}_b\rangle=
\bar u_{\cal B}[f_1\gamma_\mu+\frac{f_2}{m_{{\cal B}_b}}i\sigma_{\mu\nu}q^\nu+
\frac{f_3}{m_{{\cal B}_b}}q_\mu] u_{{\cal B}_b}\,,\nonumber\\
&&\langle {\cal B}|\bar q \gamma_\mu\gamma_5 b|{\cal B}_b\rangle=
\bar u_{\cal B}[g_1\gamma_\mu+\frac{g_2}{m_{{\cal B}_b}}i\sigma_{\mu\nu}q^\nu+
\frac{g_3}{m_{{\cal B}_b}}q_\mu]\gamma_5 u_{{\cal B}_b}\,,\nonumber\\
&&\langle {\cal B}|\bar q b|{\cal B}_b\rangle=f_S \bar u_{\cal B} u_{{\cal B}_b}\,,
\langle{\cal B}|\bar q \gamma_5 b|{\cal B}_b\rangle=f_P \bar u_{\cal B}\gamma_5 u_{{\cal B}_b}\,,
\end{eqnarray}
where $f_j$ ($g_j$) ($j=1,2,3,S$ and $P$)  are the form factors.
For the $\Lambda_b\to p$ transition,
$f_j$ and $g_j$ from different currents can be related
by the $SU(3)$ flavor and $SU(2)$ spin symmetries~\cite{Brodsky1,Chen:2008sw},
giving rise to $f_1=g_1$ and $f_{2,3}=g_{2,3}=0$.
These relations are also in accordance with the derivations from
the heavy-quark and large-energy symmetries in Ref.~{\cite{CF}}.
Note that the helicity-flip terms  of
 $f_{2,3}$ and $g_{2,3}$ vanish  due to the symmetries.
Moreover, as shown in Refs.~\cite{CF,Wei:2009np,Gutsche:2013oea},
$f_{2,3}$ $(g_{2,3})$  can only be contributed from the
loops, resulting in that they are smaller than
$f_1(g_1)$ by one order of magnitude, and can be safely ignored.
By equation of motion, we get
$f_S=[(m_{{\cal B}_b}-m_{\cal B})/(m_b-m_q)] f_1$ and
$f_P=[(m_{{\cal B}_b}+m_{\cal B})/(m_b+m_q)] g_1$.
In the double-pole momentum dependences,
$f_1$ and $g_1$ are in the forms of
\begin{eqnarray}
f_1(q^2)=\frac{C_F}{(1-q^2/m_{{\cal B}_b}^2)^2}\,,\;
g_1(q^2)=\frac{C_F}{(1-q^2/m_{{\cal B}_b}^2)^2}\,,
\end{eqnarray}
with $C_F\equiv f_1(0)=g_1(0)$.
To calculate the branching ratio of $\Lambda_b\to pM$ or $pV$, we take
the averaged decay width
$\Gamma\equiv (\Gamma_{M(V)}+\Gamma_{\bar M(\bar V)})/2$
with $\Gamma_{M(V)}$ ($\Gamma_{\bar M(\bar V)}$) for
$\Lambda_b\to p M(V)$ ($\bar \Lambda_b\to \bar p\bar M(\bar V)$).
 From Eq.~(\ref{eq1}) and Eq.~(\ref{eq2}), we can derive the ratios
\begin{eqnarray}\label{RpiK}
{\cal R}_{\pi K} &\equiv &\frac{{\cal B}(\Lambda_b\to p \pi^-)}{{\cal B}(\Lambda_b\to p K^-)}
=\frac{f_\pi^2}{f_K^2}
\frac{|\alpha_\pi|^2+|\alpha_{\bar \pi}|^2}{|\alpha_K|^2+|\alpha_{\bar K}|^2}
\frac{1+\xi^+_{\pi}}{1+\xi^+_{K}}\,,
\nonumber\\
{\cal R}_{\rho K^*}&\equiv & \frac{{\cal B}(\Lambda_b\to p \rho^-)}{{\cal B}(\Lambda_b\to p K^{*-})}
=\frac{f_\rho^2}{f_{K^*}^2 }
\frac{|\alpha_\rho|^2+|\alpha_{\bar\rho}|^2}{|\alpha_{K^*}|^2+|\alpha_{\bar K^*}|^2}\,,
\end{eqnarray}
where $\xi^+_M$ ($M=\pi,\,K$) are defined by
\begin{eqnarray}\label{Rbeta}
\xi^\pm_M\equiv
\left(\frac{|\beta_M|^2\pm |\beta_{\bar M}|^2}{|\alpha_M|^2+|\alpha_{\bar M}|^2}\right)R_{\Lambda_b\to p}\,,
\end{eqnarray}
with $R_{\Lambda_b\to p}={|\langle p|\bar u\gamma_5 b|\Lambda_b\rangle|^2}/
{|\langle p|\bar u b|\Lambda_b\rangle|^2}$, representing the uncertainty from the hadronization.
The direct CP asymmetry is defined by
\begin{eqnarray}\label{Acp}
{\cal A}_{CP}=
\frac{\Gamma_{M(V)}-\Gamma_{\bar M(\bar V)}}{\Gamma_{M(V)}+\Gamma_{\bar M(\bar V)}}\,.
\end{eqnarray}
  Explicitly, from Eqs.~(\ref{eq1}), (\ref{eq2}) and (\ref{Acp}), we obtain
  \begin{eqnarray}\label{Acp2}
{\cal A}_{CP}(\Lambda_b\to pM)&=&
\left(\frac
{|\alpha_M|^2-|\alpha_{\bar M}|^2}
{|\alpha_M|^2+|\alpha_{\bar M}|^2}+\xi^-_M\right){1\over 1+\xi^+_M}\,,\nonumber\\
{\cal A}_{CP}(\Lambda_b\to pV)&=&
\frac{|\alpha_V|^2-|\alpha_{\bar V}|^2}{|\alpha_V|^2+|\alpha_{\bar V}|^2}\,.
\end{eqnarray}

It is interesting to point out that for ${\cal R}_{\rho K^*}$ in Eq.~(\ref{RpiK}),
there is no uncertainty from  the $\Lambda_b\to p$ transition,
while both mesonic and baryonic uncertainties are totally eliminated 
for ${\cal A}_{CP}(\Lambda_b\to pV)$ in Eq.~(\ref{Acp2}).
Even for ${\cal R}_{\pi K}$ and ${\cal A}_{CP}(\Lambda_b\to pM)$,
we will demonstrate later that the hadron uncertainties can be limited.

\section{Numerical Results and Discussions }
For the numerical analysis,
the theoretical inputs of the meson decay constants and
the Wolfenstein parameters for the CKM matrix are taken
as \cite{pdg} 
\begin{eqnarray}
&&(f_\pi,f_K,f_\rho,f_{K^*})=(130.4\pm 0.2,\,156.2\pm 0.7,\,210.6\pm 0.4,\,204.7\pm 6.1)\,\text{MeV}\,,\nonumber\\
&&(\lambda,\,A,\,\rho,\,\eta)=
(0.225,\,0.814,\,0.120\pm 0.022,\,0.362\pm 0.013)\,.
\end{eqnarray}
We note that  $f_{\rho,K^*}$ are extracted from the $\tau$ decays of $\tau^-\to (\rho^-,K^{*-})\nu_\tau$,
and $V_{ub}=A\lambda^3 (\rho-i\eta)$ and
$V_{td}=A \lambda^3 (1 - i \eta - \rho)$ are used to provide the weak phase for  CP violation,
while the strong phases are coming from the effective Wilson coefficients $c^{eff}_i$ ($i=1,2,3, ...,6$).
Explicitly, at the $m_b$ scale, one has that \cite{ali}
\begin{eqnarray}
c^{eff}_1&=&1.168,\;\; c^{eff}_2=-0.365\,,\nonumber\\
10^4 \epsilon_1 c^{eff}_3&=&64.7+182.3 \epsilon_1\mp 20.2\eta - 92.6\rho +27.9\epsilon_2\nonumber\\
&&+i(44.2-16.2 \epsilon_1\mp 36.8\eta -108.6\rho + 64.4 \epsilon_2),\,\nonumber\\
10^4 \epsilon_1 c^{eff}_4&=&-194.1-329.8 \epsilon_1\pm 60.7\eta +277.8\rho -83.7\epsilon_2\nonumber\\
&&+i(-132.6+48.5 \epsilon_1\pm 110.4\eta +325.9\rho -193.3 \epsilon_2),\,\nonumber\\
10^4 \epsilon_1 c^{eff}_5&=&64.7+89.8 \epsilon_1\mp 20.2\eta - 92.6\rho +27.9\epsilon_2\nonumber\\
&&+i(44.2-16.2 \epsilon_1\mp 36.8\eta -108.6\rho + 64.4 \epsilon_2),\,\nonumber\\
10^4 \epsilon_1 c^{eff}_6&=&-194.1-466.7 \epsilon_1\pm 60.7\eta +277.8\rho -83.7\epsilon_2\nonumber\\
&&+i(-132.6+48.5 \epsilon_1\pm 110.4\eta +325.9\rho -193.3 \epsilon_2),\,
\end{eqnarray}
for the $b\to d$ ($\bar b\to \bar d$) transition, 
and
\begin{eqnarray}
&&c^{eff}_1=1.168,\,c^{eff}_2=-0.365\,,\nonumber\\
&&10^4 c^{eff}_3=241.9\pm 3.2\eta + 1.4 \rho + i(31.3\mp 1.4\eta + 3.2\rho),\,\nonumber\\
&&10^4 c^{eff}_4=-508.7 \mp 9.6\eta - 4.2\rho+ i(-93.9 \pm 4.2\eta - 9.6\rho) ,\,\nonumber\\
&&10^4 c^{eff}_5=149.4\pm 3.2\eta + 1.4\rho + i(31.3\mp 1.4\eta + 3.2\rho),\,\nonumber\\
&&10^4 c^{eff}_6=-645.5 \mp 9.6\eta- 4.2\rho + i(-93.9\pm 4.2\eta - 9.6\rho) ,\,
\end{eqnarray}
for the $b\to s$ ($\bar b\to \bar s$) transition,
where $\epsilon_1=(1-\rho)^2+\eta^2$ and $\epsilon_2=\rho^2+\eta^2$.
By adopting $C_F=0.14\pm 0.03$ from the  light-cone sum rules in Ref.~\cite{CF},
with the central value  in agreement with those in Refs.~\cite{Wei:2009np,Gutsche:2013oea},
we find that 
${\cal B}(\Lambda_b\to p K^-)=(5.1^{+2.4}_{-2.0})\times 10^{-6}$ and
${\cal B}(\Lambda_b\to p \pi^-)=(4.4^{+2.1}_{-1.7})\times 10^{-6}$,
which are consistent with the data in Eq.~(\ref{exbr}).
This is regarded to have a modest nonfactorizable effect,
as investigated by the study of $\Lambda_b\to p\pi^-$ in Ref.~\cite{CF}.
Nonetheless, since the uncertainties from the predictions exceed those of the data,
we fit $C_F$ with the data in Eq.~(\ref{exbr}), and obtain
\begin{eqnarray}\label{CF}
C_F=0.136\pm 0.009,
\end{eqnarray}
which is able to reconcile the theoretical studies of $C_F$ to the data, and to be used in our study.
Theoretical inputs in the SM for  $R_{\Lambda_b\to p}$ and $\xi^\pm_M$ in Eq.~(\ref{Rbeta})
can be evaluated, given by
\begin{eqnarray}\label{RKpi}
R_{\Lambda_b\to p}&=&1.008\,,\nonumber\\
(\xi^+_{\pi},\,\xi^+_{K})&=&(1.03\pm 0.04\pm 0.00,\,0.11\pm 0.01\pm 0.02)\,,\nonumber\\
(\xi^-_{\pi},\,\xi^-_{K})&=&(-4.0\pm 0.3\pm 0.0,\,-4.0\pm 0.2\pm 0.3)\times 10^{-3}\,,
\end{eqnarray}
where the errors for $\xi^\pm_M$ come from the CKM matrix elements and the floating $N^{eff}_c$, respectively.
We present the branching ratios and direct CP asymmetries of
$\Lambda_b\to p M(V)$ with $M(V)=K^-(K^{*-})$ and $\pi^-(\rho^-)$ in Table~\ref{pre}.

In Refs.~\cite{Aaltonen,Rosner:2014fda}, it is pointed out that
the ratio of ${\cal R}_{\pi K}$ observed by CDF~\cite{CDFexbr} or LHCb~\cite{LHCbexbr}
has not been realized theoretically, as shown in Table~\ref{BBdata}.
In particular, we note that
${\cal R}_{\pi K}=2.6^{+2.0}_{-0.5}$ in the pQCD prediction~\cite{Lu:2009cm}
is about 3 times larger than the data, but better than  other QCD calculations,
such as ${\cal R}_{\pi K}=10.7$ in Ref.~\cite{Wei:2009np}.
However, in Table~\ref{BBdata} our result of this study  shows that ${\cal R}_{\pi K}=0.84\pm 0.09$,
which agrees well with the combined experimental value of $0.84\pm 0.22$ by CDF and LHCb.
Clearly, our result  justifies the theoretical approach based on the factorization
in the two-body $\Lambda_b$ decays.
We emphasize that the ratio of ${\cal R}_{\rho K^*}$ for the vector meson modes, which is predicted to be around 4.6,
 is an interesting physical observable as it is free of the hadronic uncertainties from the baryon sectors.
A measurement for this ratio will be a firm test of the factorization approach in these baryonic decays.
\begin{table}[b]
\caption{Ratios of ${\cal R}_{\pi K}$ and ${\cal R}_{\rho K^*}$ from our calculations, the pQCD and experiments,
where the errors of our results are from the CKM matrix elements and non-factorizable effects, respectively.}
\label{BBdata}
\begin{tabular}{|c|cc|}
\hline
                                      &${\cal R}_{\pi K}$ & ${\cal R}_{\rho K^*}$ \\\hline
our result                      &$0.84\pm 0.09\pm 0.00$&$4.6\pm 0.5\pm 0.1$\\
pQCD~\cite{Lu:2009cm}&$2.6^{+2.0}_{-0.5}$&---\\
CDF~\cite{CDFexbr}~      &$0.66\pm 0.14\pm 0.08$&---\\
LHCb~\cite{LHCbexbr}  &$0.86\pm 0.08\pm 0.05$&---\\
\hline
\end{tabular}
\end{table}

As shown in Table~\ref{pre}, for the first time, the theoretical values of
${\cal B}(\Lambda_b\to p K^-)$ and ${\cal B}(\Lambda_b\to p \pi^-)$
are found to be simultaneously in agreement with the data.
Moreover, we demonstrate that the  uncertainties  from
the form factors, the CKM matrix elements and the non-factorizable effects
are small and well-controlled.

Similarly,
for the decays of $\Lambda_b\to (p K^{*-},p \rho^-)$, the predictions of
the branching ratios in Table~\ref{pre} are accessible to the experiments 
at CDF and LHCb.
Note that our results of ${\cal B}(\Lambda_b\to p K^{*-},p\rho^-)$
$\simeq (2.5,\,11.4)\times 10^{-6}$ in Table~\ref{pre}
are larger than those of $(0.3,\,6.1)\times 10^{-6}$~\cite{Wei:2009np}
and $(0.8,\,1.9)\times 10^{-6}$~\cite{Wang:2012zze} in other theoretical calculations.
\begin{table}[b]
\caption{Decay branching ratios and direct CP asymmetries of $\Lambda_b\to p M(V)$,
where the errors for ${\cal B}(\Lambda_b\to pM(V))$ arise from $f_{M(V)}$ and $f_1(g_1)$,
the CKM matrix elements and non-factorizable effects, while those
 for ${\cal A}_{CP}(\Lambda_b\to pM(V))$ are from
the CKM matrix elements and non-factorizable effects, respectively.
}\label{pre}
\begin{tabular}{|l|c|c|c|}
\hline
& our result&pQCD \cite{Lu:2009cm}
&data\\\hline
$10^{6}{\cal B}(\Lambda_b\to p K^-)$
&$4.8\pm 0.7\pm 0.1\pm 0.3$
&$2.0^{+1.0}_{-1.3}$
&$4.9\pm 0.9$ \cite{pdg}\\
$10^{6}{\cal B}(\Lambda_b\to p \pi^-)$
&$4.2\pm 0.6\pm 0.4\pm 0.2$
&$5.2^{+2.5}_{-1.9}$
&$4.1\pm 0.8$ \cite{pdg}\\
$10^{6}{\cal B}(\Lambda_b\to p K^{*-})$
&$2.5\pm 0.3\pm 0.2\pm 0.3$
&---
&---\\
$10^{6}{\cal B}(\Lambda_b\to p \rho^-)$
&$11.4\pm 1.6\pm 1.2\pm 0.6$
&---
&---\\\hline
$10^{2}{\cal A}_{CP}(\Lambda_b\to p K^-)$
&$\,\,5.8\pm 0.2\pm 0.1$
&$-5^{+26}_{-\;\;5}$
&$-10\pm 8\pm 4$ \cite{Aaltonen}\\
$10^{2}{\cal A}_{CP}(\Lambda_b\to p \pi^-)$
&$-3.9\pm 0.2\pm 0.0$
&$-31^{+43}_{-\;\;1}$
&$6\pm 7\pm 3$ \cite{Aaltonen}\\
$10^{2}{\cal A}_{CP}(\Lambda_b\to p K^{*-})$
&$19.6\pm 1.3\pm 1.0$
&---
&---\\
$10^{2}{\cal A}_{CP}(\Lambda_b\to p \rho^-)$
&$-3.7\pm 0.3\pm 0.0$
&---
&---\\\hline
\end{tabular}
\end{table}

For  CP violation,
from Eqs.~(\ref{Acp2}) and (\ref{RKpi}),
one can use the reduced forms of ${\cal A}_{CP}(\Lambda_b\to pM)\propto$
$({|\alpha_M|^2-|\alpha_{\bar M}|^2})$$/({|\alpha_M|^2+|\alpha_{\bar M}|^2})$
similar to  ${\cal A}_{CP}(\Lambda_b\to pV)$,
which indeed present the limited hadron uncertainties,
except for the factor of 1/2 for ${\cal A}_{CP}(\Lambda_b\to p \pi^-)$.
As shown in Table~\ref{pre},  our predictions of
${\cal A}_{CP}(\Lambda_b\to p \pi,\, pK^-)$ are around $(-3.9,\,5.8)$\% with
the errors less than $0.2\%$,
while the results from the data~\cite{Aaltonen} as well as the pQCD calculations are given
to be consistent with zero.

For the vector modes, as the uncertainties from the hadronizations 
have been totally eliminated  in Eq.~({\ref{Acp2}}),
we are able to obtain reliable theoretical predictions for ${\cal A}_{CP}$,
which should be helpful for experimental searches.
In particular, it is worth to note that
${\cal A}_{CP}(\Lambda_b\to p K^{*-})=(19.6\pm 1.6)\%$ is  another example of
the large and clean CP violating effects without hadronic uncertainties as
the process in the baryonic $B$ decays of $B^\pm \to K^{*\pm}\bar{p}p$~\cite{Geng:2006jt}.

Interestingly, one would ask
why the CP symmetry in $\Lambda_b\to p K^{*-}$ is  larger than those in the other baryonic decay modes.
The reason is that the term related to $a_4$
from the penguin diagram in Eq. (\ref{eq2}) can be
the primary contribution to $\Lambda_b\to p K^{*-}$ in Eq. (\ref{eq1}),
while allowing the certain contribution to the $a_1$ term from the tree diagram,
such that the apparent large interference is able to take place.
In contrast, in $\Lambda_b\to p\pi^-(\rho^-)$ and $\Lambda_b\to p K^-$,
the $a_1$  and ($a_4+r_M a_6$) terms are
dominating the branching ratios,  respectively, leaving  less rooms for the interferences.
Clearly, ${\cal A}_{CP}(\Lambda_b\to p K^{*-})$ as well as the CPAs in other modes
should receive more attentions, which have also been emphasized in
Ref.~\cite{Gronau:2013mza}.
Finally, we  remark that our approach can be extended
to the two-body decay modes of other beauty baryons (${\cal B}_b$),
such as $\Xi_b$, $\Sigma_b$ and $\Omega_b$.

\section{Conclusions}
Based on the generalized factorization method and $SU(3)$ flavor and $SU(2)$ spin symmetries,
we have simultaneously explained the recent observed decay branching ratios in $\Lambda_b\to p K^-$ and
$\Lambda_b\to p \pi^-$  and obtained the ratio of ${\cal R}_{\pi K}$ being $0.84\pm 0.09$, which agrees well 
with the combined experimental value 
of $0.84\pm 0.22$ from CDF and LHCb,
demonstrating a reliable theoretical approach
to study the two-body $\Lambda_b$ decays.
We have also predicted that
${\cal A}_{CP}(\Lambda_b\to p K^-)=(5.8\pm 0.2)\%$ and
${\cal A}_{CP}(\Lambda_b\to p \pi^-)=(-3.9\pm 0.2)\%$
with  well-controlled uncertainties,
whereas the current data for these CPAs are consistent with zero.
We have used this approach to study the corresponding vector modes.
Explicitly, we have found that
${\cal B}(\Lambda_b\to p K^{*-},\, p \rho^-)=(2.5\pm0.5,\,11.4\pm2.1)\times 10^{-6}$
with  ${\cal R}_{\rho K^*}=4.6\pm0.5$
and ${\cal A}_{CP}(\Lambda_b\to p K^{*-},\,p \rho^-) =(19.6\pm1.6,\, -3.7\pm0.3)\%$.
Since our prediction for ${\cal A}_{CP}(\Lambda_b\to p K^{*-})$
is large and free of both mesonic and baryonic uncertainties from the hadron sector, it would be
the most promised direct CP asymmetry
to be measured by the experiments at  CDF and LHCb
to test the SM and search for some possible new physics.

\section*{ACKNOWLEDGMENTS}
This work was partially supported by National Center for Theoretical
Sciences,  National Science Council
(NSC-101-2112-M-007-006-MY3) and National Tsing Hua
University~(103N2724E1).

\end{document}